\documentclass[twocolumn]{aastex631}

\usepackage{amsmath}

\graphicspath{{./}{figs/}}

\begin{document}

\title{Flares, jets and quasi-periodic outbursts from neutron star merger remnants}

\author[0000-0002-0491-1210]{Elias R. Most}
\correspondingauthor{Elias R. Most}
\email{emost@princeton.edu}
\affiliation{Princeton Center for Theoretical Science, Jadwin Hall, Princeton University, Princeton, NJ 08544, USA}
\affiliation{Princeton Gravity Initiative, Jadwin Hall, Princeton University, Princeton, NJ 08544, USA}
\affiliation{School of Natural Sciences, Institute for Advanced Study, 1 Einstein Drive, Princeton, NJ 08540, USA}
\author[0000-0001-9185-5044]{Eliot Quataert}
\affiliation{Department of Astrophysical Sciences, Princeton University, Princeton, NJ 08544, USA}
\affiliation{Princeton Center for Theoretical Science, Jadwin Hall, Princeton University, Princeton, NJ 08544, USA}

\begin{abstract}
Using numerical relativity simulations with a subgrid dynamo prescription to generate strong initial magnetic fields, we investigate the possibility of launching a jet-like outflow from the hypermassive neutron star (HMNS) during the early stages of the merger, prior to the remnants collapse to a black hole.
We demonstrate that buoyant instabilities in the strongly magnetized HMNS can lead to a periodic emission of powerful electromagnetic flares shortly after the merger. These are followed by a collimated mildly relativistic outflow. Both types of outflows feature quasi-periodic kilohertz substructure.  These early-time outflows may power precursors to short-duration gamma-ray bursts (SGRB) or in some cases the entire SGRB.   While the overall temporal power spectrum we find broadly agrees with the one recently reported for quasi-periodic oscillations in the SGRB GRB910711, our simulations suggest that the 
periodic electromagnetic substructure is {dominated} by magnetohydrodynamic {shearing} processes rather than correlating with the corresponding post-merger gravitational wave signal.
\end{abstract}


\section{Introduction}

Binary neutron star mergers can feature some of the most extreme magnetic
field configurations in the universe (see, e.g., \citealt{Baiotti:2016qnr}
for a review).
Although magnetic field strengths in the inspiralling systems may be small (and
dynamically unimportant), turbulent small-scale process can potentially amplify the
magnetic field during merger \citep{Price:2006fi,Kiuchi:2015sga}, reaching dynamically important field strengths
\citep{Kiuchi:2017zzg,Aguilera-Miret:2021fre}. 
Despite their importance, the magnetic field topology after merger remains
largely unconstrained.\\
While several numerical relativity simulations of magnetized binary neutron
star mergers have been performed (e.g.,
\citealt{Kiuchi:2014hja,Dionysopoulou:2015tda,Palenzuela:2015dqa,Ciolfi:2017uak,Most:2019kfe,Palenzuela:2021gdo,Werneck:2022exo,Combi:2022nhg,Kiuchi:2022nin}),
turbulent magnetic field amplification during the collision
\citep{Price:2006fi,Kiuchi:2015sga,Aguilera-Miret:2020dhz}, which likely
dominates the post-merger magnetic field geometry
\citep{Aguilera-Miret:2021fre}, is not fully captured by current
simulations \citep{Kiuchi:2017zzg}. 
Although magnetic fields in the inspiral are not likely to strongly
affect the orbital dynamics or gravitational wave emission \citep{Ioka:2000yb,Zhu:2020imp}, they could produce electromagnetic precursors prior to the merger
(see,
e.g., \citealt{Hansen:2000am,Tsang:2011ad,Palenzuela:2013hu,Carrasco:2020sxg,Most:2020ami,Zhang:2020eou,Most:2022ayk}).  In addition, magnetic fields undoubtedly
play a major role in the dynamics of the post-merger evolution and
its connection to electromagnetic afterglows and the production of
gamma-ray bursts (GRBs)  (see respectively, e.g., \citealt{Metzger:2019zeh} and \citealt{Berger:2013jza}
for recent reviews). In particular, magnetic fields might 
drive winds from the (meta-)stable remnant star formed in some mergers
\citep{Metzger:2018qfl,Ciolfi:2020wfx} and facilitate most of the mass ejection from post-merger accretion
disks (e.g., \citealt{Fernandez:2013tya,Martin:2015hxa,Siegel:2017nub,Fernandez:2018kax}). 
In regimes in which a remnant neutron star is  rotationally supported
against gravitational collapse, magnetic fields provide various means of
transporting angular momentum, critically affecting the lifetime of the
system \citep{Gill:2019bvq,Margalit:2022rde}.
They also enable jet launching ultimately resulting in (short) GRBs
\citep{Rezzolla:2011da,Ruiz:2016rai}. \\

Over the past years, the connection between binary neutron star mergers and GRBs has been confirmed directly
by the coincident detection of the short GRB GRB170817A with the
gravitational-wave (GW) event GW170817 \citep{LIGOScientific:2017zic}.
Adding to this picture, 
\citet{Chirenti:2023dzl} recently reported the discovery of kilohertz
quasi-periodic oscillations (QPOs) in a small subset of previously detected
sGRBs.  Assuming a neutron star merger as the origin of these sGRBs, one
possible place to look for kilohertz QPOs are the oscillations of a
(meta-)stable hypermassive neutron star (HMNS) formed during the collision
\citep{Chirenti:2019sxw}. These have been shown to be quasi-universally
related to properties of the inspiralling neutron stars and the dense
matter they contain (e.g.,
\citet{Bauswein:2012ya,Bauswein:2011tp,Takami:2014zpa,Bernuzzi:2015rla,Raithel:2022orm}).
If this association could be made, it would potentially open up a new way
of placing constraints on the dense matter equation of state (e.g.,
\citealt{Ozel:2016oaf}), provided the redshift of these events could be
determined \citep{Chirenti:2023dzl}.

Despite these exciting prospects,
there are a number of questions about the connection between the relativistic outflow that powers the gamma-ray emission and the post-merger oscillations. Post-merger
oscillations are most pronounced in the first few milliseconds after the
collisions (see, e.g., \citealt{Nedora:2021eoj,Most:2021zvc} for recent
discussions). However, launching a relativistic outflow or jet from a
stellar merger typically happens on longer timescales
\citep{Siegel:2014ita,Ciolfi:2020hgg}, when these oscillations have long
decayed away. Therefore, launching a relativistic jet-like outflow, that
could contain imprints of the oscillating HMNS at its base would require
establishing a strongly collimated magnetic funnel region immediately after
merger \citep{Shibata:2011fj,Mosta:2020hlh}.
Due to the difficulty capturing small-scale dynamo physics in present simulations, such a scenario can currently only be realized in numerical simulations with subgrid dynamo prescriptions (e.g., \citealt{Tomei:2019zpj,Shibata:2021xmo}).\\ 

We here present binary neutron star merger simulations, which are augmented
with a sub-grid dynamo prescription enabling us to freely adjust the
magnetic field strengths produced during the merger (see Sec.
\ref{sec:methods}). Using this idealized approach, we provide a
proof-of-principle demonstration that for extreme magnetic field strengths
in excess of $B \simeq 10^{17}\, \rm G$, HMNS can quasi-periodically emit
flares (see Sec. \ref{sec:flares}) and relativistic outflows (see Sec.
\ref{sec:outflows}), which feature kilohertz substructures (see Sec.
\ref{sec:qpo}). Despite the caveats of using such an effective approach,
our work clearly demonstrates the importance of magnetic field
amplification and the resulting post-merger topology in understanding sGRB
production, particularly in the earliest post-merger phases.

\begin{figure}
    \centering
    \includegraphics[width=0.45\textwidth]{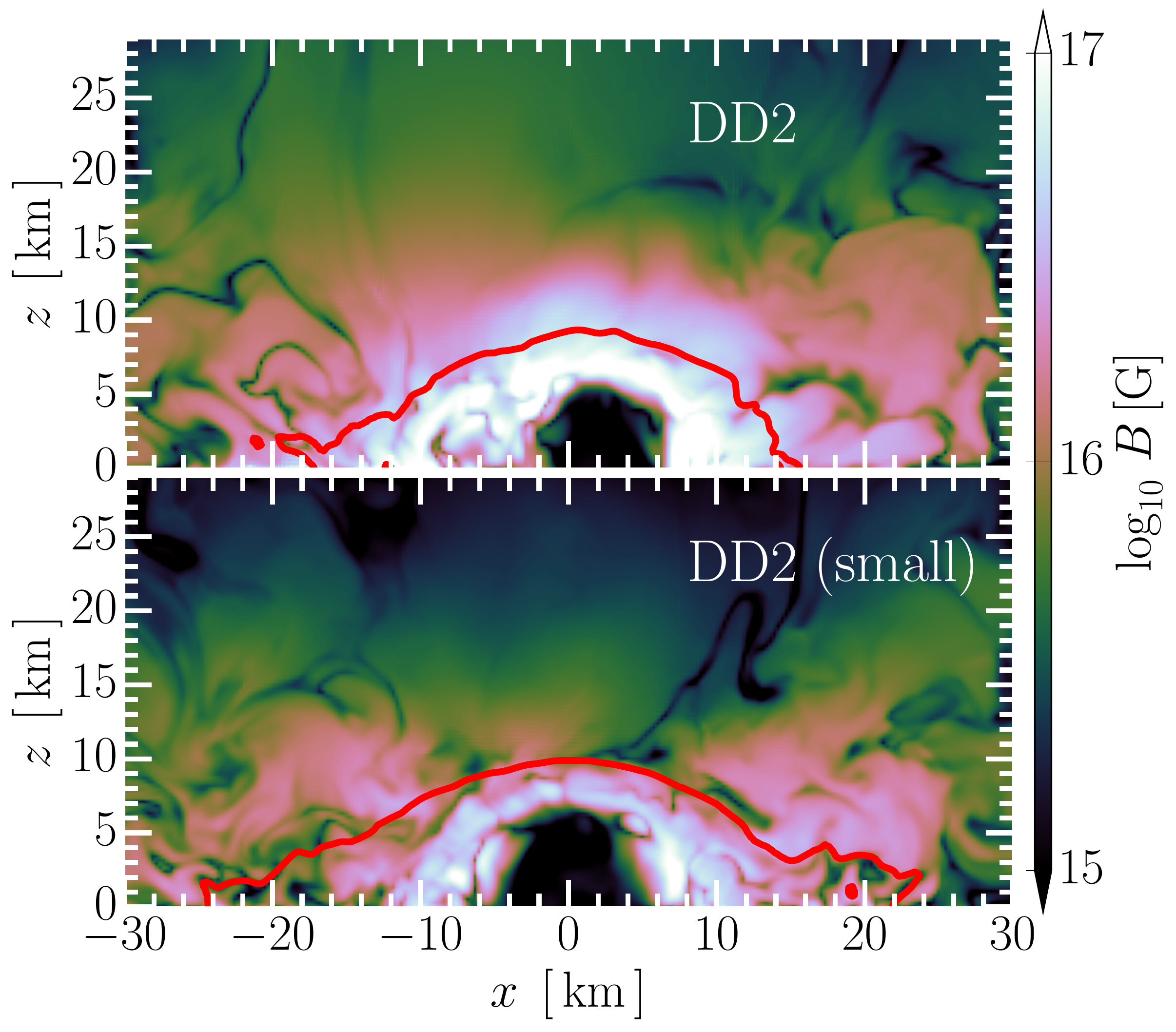}
    \caption{Magnetic field strength, $B$, in the hypermassive neutron star around the time of the onset of flaring. The solid red line approximately marks the surface of the star. The field close to the surface has been amplified to high field strengths due the use of a subgrid dynamo model, mimicking the effect of small scale field amplification during merger. Shown are high (top) and low (bottom) magnetization cases.}
    \label{fig:B_initial}
\end{figure}
\begin{figure*}
    \centering
    \includegraphics[width=0.85\textwidth]{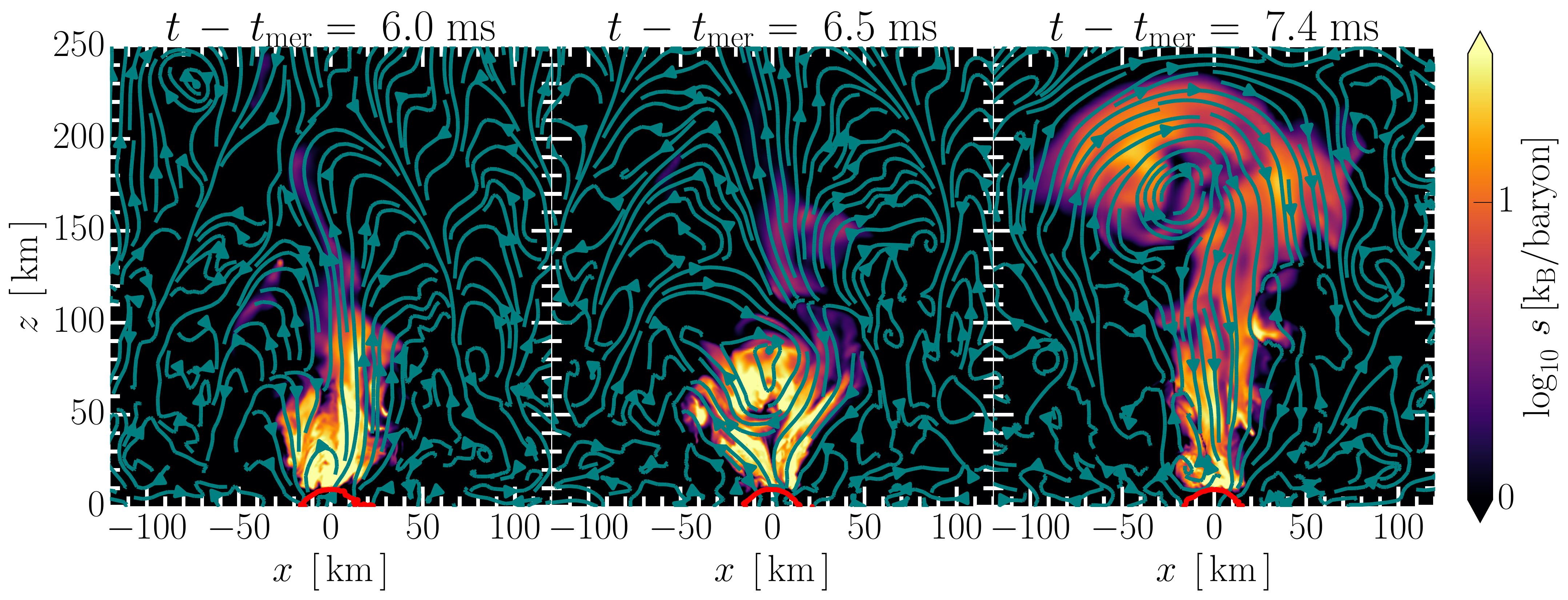}
    \caption{Flaring process at the surface (red) of the neutron star merger remnant. Show in color are the magnetic field, the entropy per baryon $s$ (indicating heating inside the flare). The times $t$ are stated relative to the time of merger $t_{\rm mer}$.
    {(\it Left)} Built-up of magnetic stresses due to buoyancy of field
    lines from the star. {(\it Center)} Inflation the connected flux tube
    due to strong differential rotation at the surface of the star, where
    both footpoints are anchored. {(\it Right)} Detachment of
  the flare within $\simeq 1 \rm ms$ of its creation. In our simulations, this cycle repeats
quasi-periodically until a steady outflow is launched.}
    \label{fig:Bstar}
\end{figure*}

\section{Methods}\label{sec:methods}
We perform numerical simulations of neutron star coalescence
in full general relativity using the \texttt{Frankfurt/IllinoisGRMHD (FIL)}
code \citep{Most:2019kfe,Etienne:2015cea}.  
\texttt{FIL} solves the equations of general-relativistic
magnetohydrodynamics (GRMHD) in dynamical spacetimes \citep{Duez:2005sf}, evolved
using the Z4c formulation \citep{Hilditch:2012fp}.  
Furthermore, the code makes use of high-order numerical methods allowing us
to capture especially magnetic field dynamics accurately at lower numerical
resolutions \citep{DelZanna:2007pk}.
Following the GRMHD benchmark runs presented in \citep{Most:2019kfe}, we perform
simulations on a mesh-refined grid with the highest numerical resolution of $\Delta x=250\, \rm m$.  We further
make use of a neutrino leakage scheme \citep{Ruffert:1995fs}, although neutrino aspects are not important for this work. 
We adopt the DD2 equation of state having a radius of $R_{1.4} = 13.2\,\rm km$ \citep{Hempel:2009mc},
as well as the APR$_{\rm LDP}$ equation of state with $R_{1.4} = 12.2\, \rm km$ for comparison \citep{Schneider:2019vdm}.
We further consider a GW170817-like configuration with a chirp mass
$\mathcal{M}=1.186\, M_\odot$ and a mass ratio $q=0.85$ at an initial
separation of $45\, \rm km$. The initial data is computed using
\texttt{LORENE} \footnote{\texttt{https://lorene.obspm.fr/}} (see also, e.g., \citealt{Papenfort:2021hod}). 
Furthermore, we start with an initial poloidal
magnetic field in the pre-merger stars given by the toroidal vector potential
$A_\phi = A_0 \max (0, \left(p-10^{-3} p_{\rm max} \right)^2
)$, where $p$ is the pressure and $p_{\rm max}$ its maximum value
inside the inspiraling stars \citep{Liu:2008xy}. The constant $A_0$ is chosen such
as to obtain field strengths of around $10^{15}\, \rm G$.
More details on the setup can be found in
\citet{Most:2019kfe} and \citet{Most:2021ktk}.\\
\begin{figure*}
  \centering
  \includegraphics[width=\textwidth]{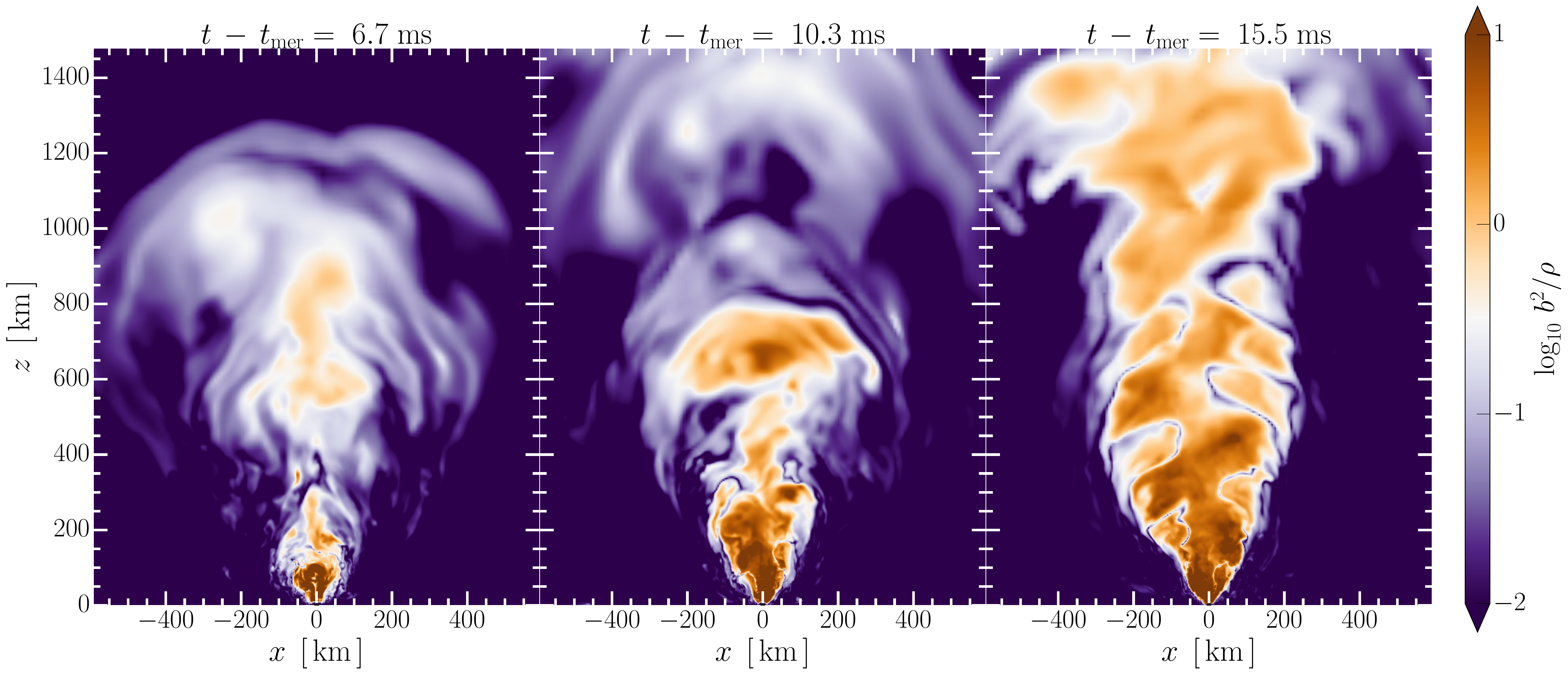}
  \caption{Propagation of the flares away from the merger site. Several flaring episodes are visible in the center pannel. Shown in
  color is the magnetization $\sigma=b^2/\rho$, where $b^2$ is the comoving
magnetic field energy density, and $\rho$ the baryon rest-mass density. All times, $t$, are stated relative to the time of merger, $t_{\rm mer}$.}
  \label{fig:b2rho}
\end{figure*}
{\it {Subgrid dynamo model}.} Even with the use of high-order methods, it is
currently very difficult to numerically capture the appearance of
magneto-turbulence generating most of the magnetic field strengths present
in the post-merger phase \citep{Price:2006fi,Kiuchi:2015sga}, as this would require at least
$10\times$ finer resolutions than applied in this study
\citep{Kiuchi:2017zzg,Aguilera-Miret:2020dhz}. Since
the mechanism presented here heavily relies on the presence of very strong
field strengths near the surface of the star formed in the merger, we opt
to drive additional magnetic field amplification by means of an effective
sub-grid dynamo model. More specifically, we have extended \texttt{FIL}
to explicitly model the mean-field dynamo $\alpha$-effect (e.g., \citet{Gruzinov:1994zz,1995ApJ...449..739B}), which in our case
corresponds to the assumption that the comoving electric field $e^\mu =
\kappa b^\mu$ is proportional to the comoving magnetic field $b^\mu$
\citep{Bucciantini:2012sm}. We
dynamically adjust $\kappa$ to vary between $\kappa_{\rm min} = 0$ and $\kappa_{\rm max}=0.05$ such that the effective magnetizations in the
outer parts of the star are driven to magnetization ratios of $\sigma = b^2/\rho \sim
0.01 $ (corresponding to magnetic field strengths in excess of $B\gtrsim 10^{17}$\, \rm G),
where $\rho$ is the baryon rest-mass density, and deactivate the dynamo
whenever this magnetization is reached or exceeded.  We also perform a comparison run with a target $\sigma= 0.001$. 
The resulting magnetic field configuration prior to the onset of flaring is shown in Fig. \ref{fig:B_initial}.
These values of $\kappa$ have been chosen to be consistent with the amplification timescale of Kelvin-Helmholtz dynamos seen in high-resolution simulations \citep{Kiuchi:2014hja,Kiuchi:2017zzg,Palenzuela:2021gdo,Chabanov:2022twz}. While appropriate for an initial demonstration of the effects considered here, more detailed follow-up work on mean-field dynamo models for neutron star mergers will be needed, similar to progress made in adjacent fields (e.g., \citealt{Gressel:2015mxa,Hogg:2018zon,White:2021erz,liu2022subgrid}).\\

The field strengths driven by the subgrid dynamo  do not by themselves triggger outbursts; their critical role is to drive the star to a regime
where kinematic dynamo process operating on macroscopic scales can take
over to drive further field growth.  Rayleigh-Taylor instabilities present during merger \citep{Skoutnev:2021chg} may also provide field amplification not captured in global simulations. The subgrid prescription is employed here merely to facilitate a controlled experiment on the role of strong post-merger magnetic fields, allowing us to drive flares and relativistic outflows in the early post-merger evolution.
More details will be presented in a forthcoming work.
\section{Results} 
In the following, we report on our findings of
flares and quasi-periodic relativistic outbursts from an ultra-magnetized
binary neutron star merger remnant. We proceed in three steps: First, we
will detail how a strongly magnetic outflow forms. Second, we will analyze
its properties, and third, we will compare the quasi-periodic substructure
in the electromagnetic outbursts to those in the gravitational wave
signal.

\subsection{Flares from the early post-merger stage}\label{sec:flares}
Shortly after the merger, small-scale turbulence will amplify the magnetic
field throughout the merger remnant \citep{Price:2006fi,Aguilera-Miret:2020dhz}, which can be predominantly
driven by the Kelvin-Helmholtz instability \citep{Kiuchi:2015sga}, or potentially also by
small-scale Rayleigh-Taylor instabilities \citep{Skoutnev:2021chg}, 
where the dependence on the pre-merger field configuration might be weak \citep{Aguilera-Miret:2021fre}.
In order to provide a simple proof-of-principle calculation, we mimick the
effects of this process by using a mean field $\alpha$-dynamo model (see
Sec. \ref{sec:methods}). Since the merger will result in a predominantly
toroidal magnetic field topology \citep{Giacomazzo:2010bx}, the $\alpha-$effect will serve
to convert part of this toroidal field into poloidal field \citep{1978RSPTA.289..459A}. Enhanced
amounts of poloidal field are also seen at ultra-resolution simulations
that begin to resolve small-scale turbulence \citep{Kiuchi:2017zzg}. This leaves us in a
situation after the merger where the magnetic field outside of the newly
formed remnant is dynamically unimportant, whereas inside and close to the
surface it is very strong (see also Fig. \ref{fig:B_initial}).
We then observe the appearance of buoyant
instabilities near the surface \citep{1978RSPTA.289..459A}, leading the poloidal components to effectively
buckle up and leak out of the star. Such a behavior has been numerically investigated
in detail in the case of isolated strongly magnetized neutron star by
\citet{Kiuchi:2011yt}.  
We demonstrate this situation in the left panel of Fig.
\ref{fig:Bstar}.  Here close to the stellar surface a connected loop sticking
out of the star has formed. This loop is strongly magnetically
dominated and the low-density plasma will get heated, allowing us to use the specific entropy,$s$,
as an effective tracer of the dynamics of this loop.  Since HMNS formed 
in mergers are strongly differently rotating \citep{Hanauske:2016gia,Kastaun:2016yaf},
the two footpoints of this extruding loop anchored at different latitudes
will experience a net shearing motion.  Crucially, this will build up a twist in this
loop causing it to inflate (see center panel of Fig. \ref{fig:Bstar}).\\  This
situation has been studied extensively in the case of magnetars \citep{Parfrey:2013gza,Carrasco:2019aas,Mahlmann:2023ipm,Sharma:2023cxc}
and even for pre-merger magnetospheres of merging neutron stars \citep{Most:2020ami,Most:2022ojl}.
A very similar scenario to the one presented here has also been worked out for the case of white dwarf mergers leading to a stable differentially rotating white dwarf remnant \citep{Beloborodov:2013kpa}.
In fact, it has been shown that as the twist is increased the inflated
bubble will eventually detach and be ejected from the system (see
right panel of Fig. \ref{fig:Bstar}). This will be triggered roughly after
a 180$^\circ$ twist of the connected flux bundle \citep{Parfrey:2013gza}. Different from the
magnetar case, this mechanism operates at a time when post-merger
oscillations are strongest, so that reconnection can be partially sped up
by forcing it due to radial compression of the connecting current sheet
between the foot points. Since this compression correlates directly with
the radial motion of the stars \citep{Stergioulas:2011gd}, we naively might expect the flare timing
to be correlated with the post-merger oscillations; we will study this explicitly in Sec. \ref{sec:qpo}.

As the connected flux tubes erupt they will produce
open field lines, which will not necessarily reconnect into closed loops
again. This will gradually result in the formation of an open field line
structure, a requirement for launching collimated (jet-like) relativistic outflows.  
In fact, our simulations show the appearance of multiple flaring episodes
before a roughly constant strongly magnetized outflow is observed. In that
sense, our simulations demonstrate how (meta-)stable neutron star remnants can
naturally produce conditions favorable to jet launching. As such,
they complement previous works, where large scale coherent field
structures where manually superimposed on the post-merger system \citep{Shibata:2011fj,Siegel:2014ita,Mosta:2020hlh}. 
\begin{figure}
  \centering
   \includegraphics[width=0.5\textwidth]{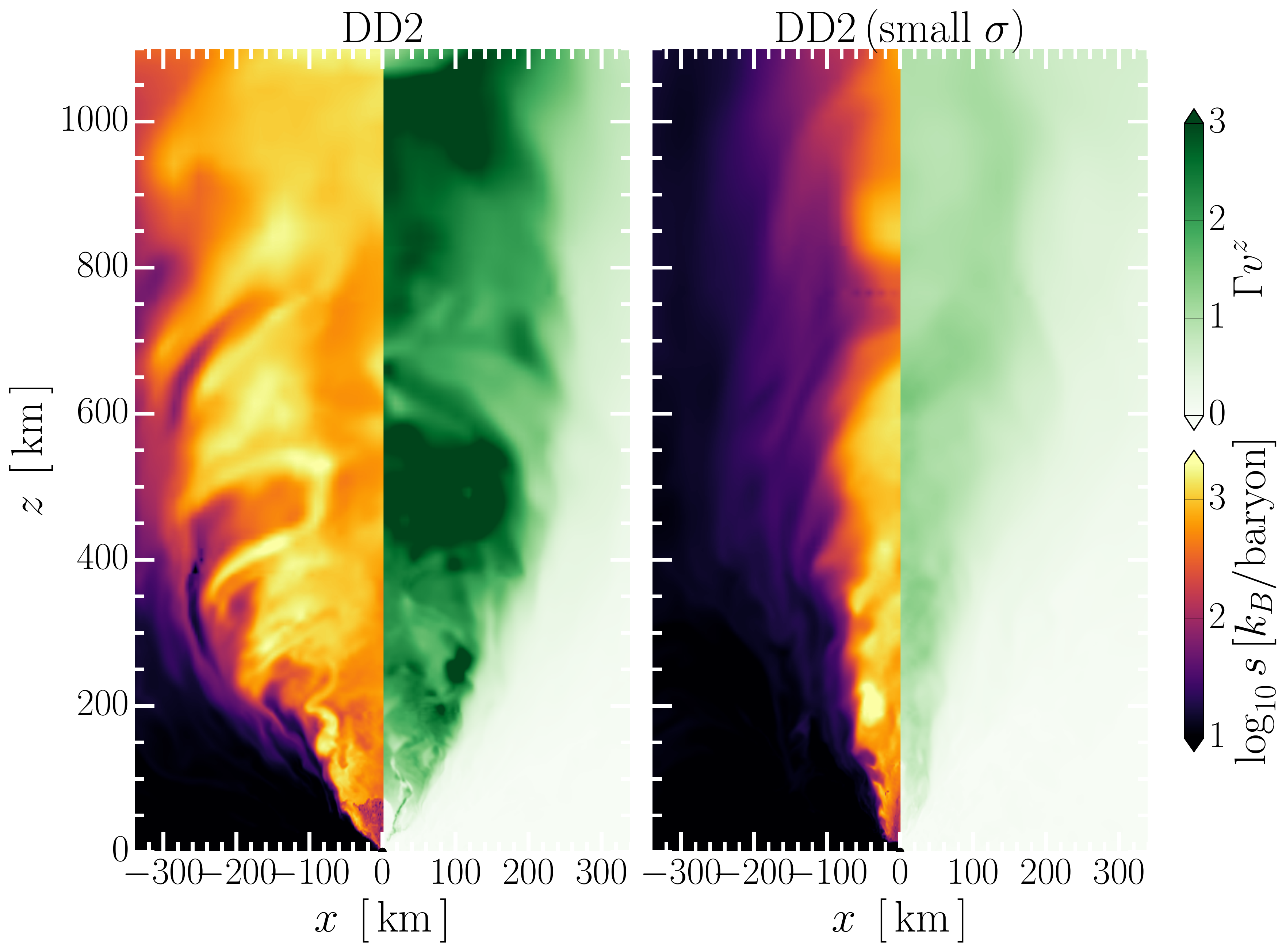}
  \caption{Same as Fig. \ref{fig:b2rho} (right panel), but showing the entropy $s$ per baryon and vertical velocity component $\Gamma v^z$ of the outflow. Shown are target magnetizations $\sigma = 0.01$ (left) and $\sigma=0.001$ (right) for the models using the DD2 equation of state. A distinct quasi-periodic substructure is visible in the mildly relativistic outflows for both models.}
  \label{fig:s_Gamma}
\end{figure}
\subsection{Electromagnetic outbursts and jet-like outflows}\label{sec:outflows}
We next focus on the propagation of the flares and relativistic outflows.
In Fig. \ref{fig:b2rho}, we track the evolution of the flares on scales of $~1,400\, \rm km$ from the star.
We focus explicitly on the high target magnetization DD2 case, although the behavior is qualitatively similar in all cases.

Starting at early times (left panel of Fig. \ref{fig:b2rho}), we can see that a strongly magnetized region, $b^2 \gg \rho$, is
building up close to the star, which gives rise to periodic flaring episodes (see also Fig. \ref{fig:Bstar}).
We can clearly identify multiple flaring episodes (white regions at distances $>800\, \rm km$).
This is most prominently shown in the middle panel of Fig. \ref{fig:b2rho}, where a compressed pancake-shaped
magnetically dominated bubble is visible, which propagates outwards with Lorentz factors $\Gamma >2$.\\
The continued emission of flares produces open field lines which ultimately pave the wave for driving 
a continued relativistic outflow.
At times $>15\, \rm ms$ after merger (see Fig. \ref{fig:b2rho}, right panel), we can indeed see that a sustained magnetically dominated relativistic outflow sets in.
The flow structure is helical (in line with, e.g., \citet{Ciolfi:2020hgg}), and features a distinct substructure. This structure roughly resembles internal kink instabilities seen in magnetic tower jets launched from rotating proto-neutron stars \citep{Bromberg:2015wra}.
In order to capture the properties of the final outflow pattern more quantitatively, Fig. \ref{fig:s_Gamma} shows
the relativistic vertical velocity component $\Gamma v^z$ for both DD2 cases, where $\Gamma$ is the Lorentz factor. 
\begin{figure}
  \centering
  \includegraphics[width=0.48\textwidth]{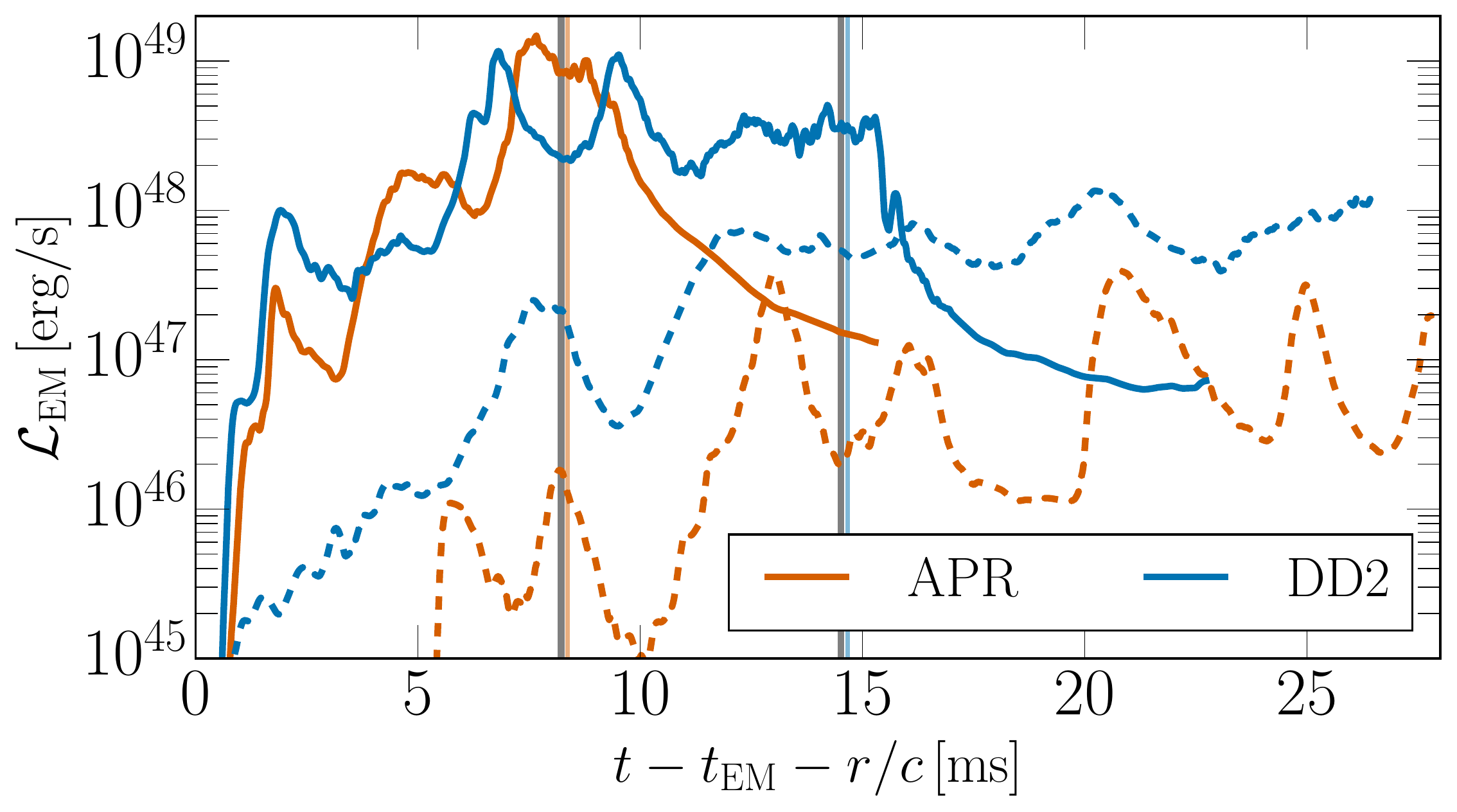}
  \caption{Electromagnetic luminosity $\mathcal{L}_{\rm EM}$ (i.e., Poynting flux) relative to the time of first electromagnetic outburst, $t_{\rm EM}$, extracted at a radius r=$236\, \rm km$ from the stellar remnant. Solid lines refer to high magnetization models, $\sigma=0.01$, dashed lines to low magnetization models, $\sigma = 0.001$. Times, $t$, are stated relative to the time $t_{\rm EM}$ of the first flare launch. For the high magnetization cases, black hole formation is indicated by a vertical black line.}
  \label{fig:LEM}
\end{figure}
We can see that low-density material inside the outflow has been strongly heated up as indicated by the
entropies, $s$, per baryon. This also allows us to precisely track the location of the flares as they move outwards, as seen in Fig. \ref{fig:Bstar}.
Indeed, we can identify distinct substructures of low entropy -- high velocity regions and vice-versa.
In the high-magnetization target case (left panel), we find that there are distinct bursts with
Lorentz factors $\Gamma >3$, consistent with similar flares launched in pre-merger
magnetospheres \citep{Most:2022ayk}. 
In between the flares we observe sub-bursts at Lorentz factors $\Gamma \simeq 2$, providing a distinct time variable substructure to the outflow.
These flares have the potential to further accelerate as they propagate outwards, 
with their evolution likely being similar to the boosted fireball sGRB model of \citet{Duffell:2013wra}, as well as the fireball model for white dwarf mergers \citep{Beloborodov:2013kpa}. 
While our simulations can capture the initial launching of flares and jet-like outflows, the later evolution will be dominated by the density of the environment. Our simulations require the use of a (constant) density floor $\rho_{\rm atm} \simeq \times 10^5 \rm g/cm^3$ (see, e.g., \citealt{Poudel:2020fte} for a recent discussion). At large distances($>1000\, \rm km$) this becomes comparable to densities in the flares and outflows and will contaminate them, causing $b^2/\rho$ to drop. Separate simulations of the late-time propagation of flares and outflows will be required to clarify the exact magnetization, and environmental impact of the outflows described here.

We next investigate how these findings depend on the initial magnetic field amplification. To this end, we performed simulations with lower target $\sigma=0.001$ for the same binary configurations. For these models (right panel of Fig. \ref{fig:s_Gamma}), we find a less broad and more collimated outflow structure.
Collimation is provided by the disk and stellar winds.
The flow in this case still features a clear time-variable substructure as can be seen by gradients in the specific entropy $s$.
Different from the high magnetization target case, the vertical velocity is lower with $\Gamma <2$, much more in line 
with incipient jet-like outflows found in other simulations \citep{Ruiz:2016rai,Mosta:2020hlh}. \\

Having described the flaring aspects of the outflow, we now want to provide a quantitative aspect of the electromagnetic luminosity of the outflow, for both flares and jets. 
We now quantify the outflow and jet power in terms of their Poynting luminosity, $\mathcal{L}_{\rm EM} = \oint_{r=\rm const} T_{{\rm
  EM}\,r}^t$, which we extract at a constant radius $r\simeq 236\, \rm km$
from the HMNS. Here, $T_{\rm EM}^{\mu\nu}$ is the electromagnetic stress-energy tensor \citep{Duez:2005sf}. This provides an upper bound for the actually observed luminosity. Quantifying the latter would require a more detailed investigation of the emission mechanism, not included in our simulations. Since most of the outflow is magnetically dominated, the Poynting flux will contain the bulk energy contribution of the outflows (although there is a matter component as well; e.g., \cite{Shibata:2011fj}).  We stress that this is appropriate since the entire polar region, including the flares, are magnetically dominated. The resulting evolution is shown in
Fig. \ref{fig:LEM} for all four models, with times stated relative to the
onset of the electromagnetic outburst $t_{\rm mer}$.\\ Focusing on the strong
amplification models (solid curves), we can see that soon after merger the
electromagnetic luminosity rises strongly. Initially at $t-t_{\rm EM} - r
= 2\, \rm ms$ the luminosity is dominated by
magnetized mass debris ejected during the collision, but decays quickly
over a time scale of one millisecond. At this point, driven by the early
emission of flares, a strong outflow develops, increasing the luminosity by
about one order of magnitude. We can see that this outflow is periodic,
with a dominant period of $2-4\, \rm ms$, depending on the model. We can
also see, that a sub-dominant quasi-periodic structure is present, but at
significantly lower power. The overall evolution is consistent between
the models. 

Due to increased magnetic braking \citep{Shapiro:2000zh}, the remnant {approaches solid body rotation} rather rapidly, triggering an early collapse of both models within about $15\, \rm ms$
from the onset of flaring. 
Due to this shortened lifetime of the HMNS, the number of quasi-periodic outbursts prior to black hole formation is lower than it otherwise might be. This is especially the case for the potentially more luminous high-magnetization simulations, which collapse to a black hole more quickly.
Of course, the time to collapse also depends on the initial masses of the inspiraling binary and the underlying equation of state  (e.g., \citealt{Hotokezaka:2011dh,Hotokezaka:2013iia}). For this study, these have been chosen to be roughly consistent with the GW170817-event \citep{LIGOScientific:2017zic}, although there is no indication that these would be preferred parameters for the scenario considered here. On the contrary,  for a lower mass merger, we would expect {angular momentum redistribution} to lead to a stable uniformly rotating star (see also \citet{Bozzola:2017qbu,Weih:2017mcw}), prolonging the duration of the quasi-periodic eruption phase and likely increasing potential detection prospects. Once the remnant is uniformly rotating, further twist will not build up efficiently in the magnetic field, and we expect to 
launch a collimated jet-like outflow \citep{Mosta:2020hlh}. The EM power in that jet could significantly exceed that found here, which is a small fraction of the maximal spin-down power of a near-breakup HMNS.\\

For our low magnetization cases (dashed lines in Fig. \ref{fig:LEM}), the star powers an outflow at roughly constant luminosity \citep{Shibata:2011fj}.
\begin{figure}
  \centering
     \includegraphics[width=0.48\textwidth]{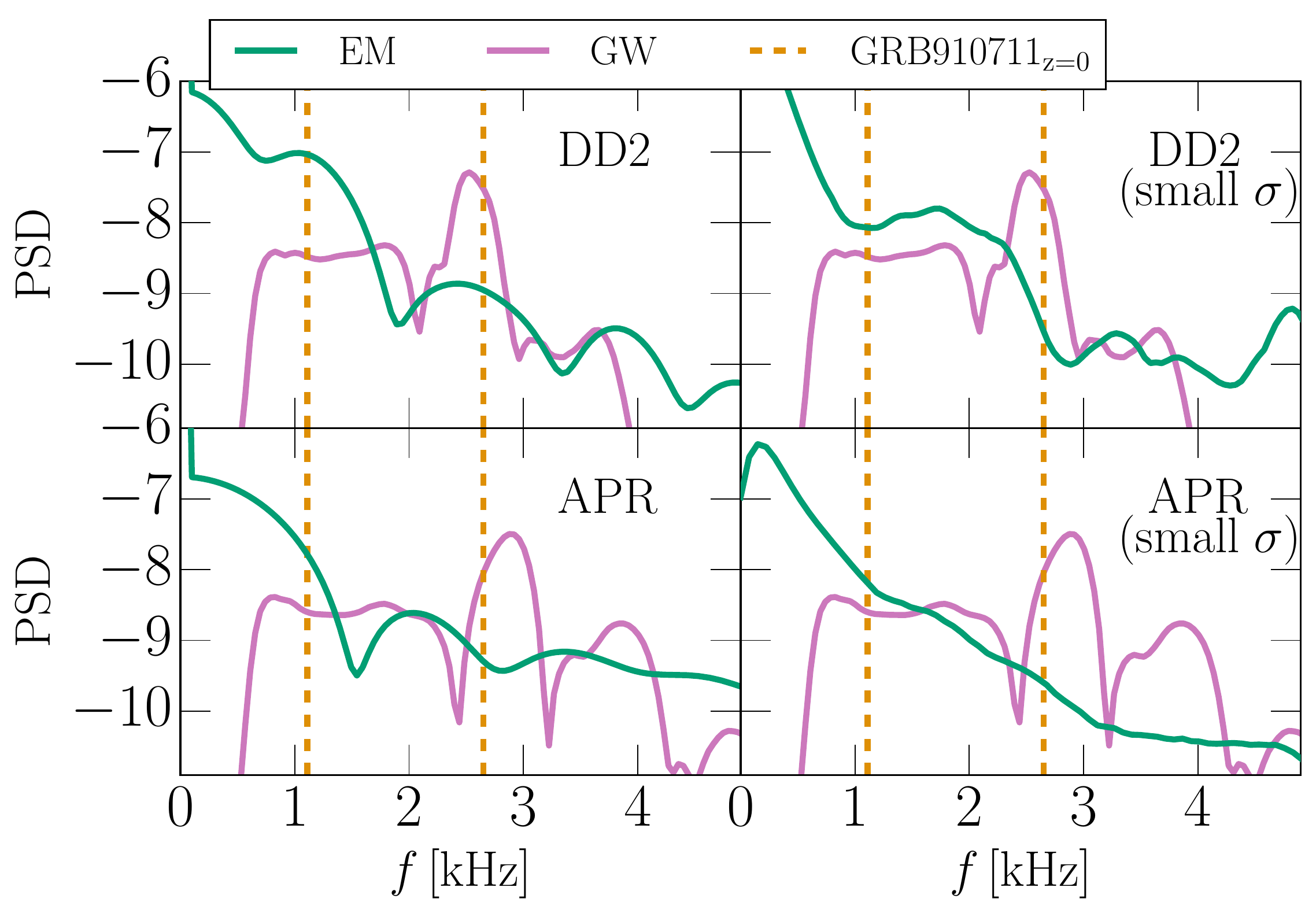}
  \caption{Frequency spectrum (logarithmic power spectral density (PSD)) of the electromagnetic (EM) and gravitational waves (GW) luminosities. Note that the units of the PSDs have been rescaled and shifted relative to each other to aid a direct comparison. }
  \label{fig:freq}
\end{figure}
In these cases, 
we can see that initially, the outflow grows at a much slower
exponential rate.  This is consistent with Parker-type buoyant
instabilities, expected for these configurations to require $b^2 \simeq
P$, in the unstable regions, now building up much slower than in the more strongly magnetized cases \citep{Kiuchi:2011yt}. 
Quasi-periodic flaring is present, but at decreased power and longer time intervals compared to the strongly magnetized cases.
At late times a steady, but modulated outflow reaches luminosities of about $10^{48}\,\rm erg/s$.\\
In fact, since the sub-grid dynamo action switches off at lower magnetizations, this
slower exponential growth is driven entirely by the ideal MHD sector.
This strengthens our result in that the final launching of the flares
likely does not depend on the dynamo itself, but rather on the presence of
ultra-strong magnetic fields (see also \citet{Kiuchi:2011yt}).\\

Since our simulations do not capture the effects of neutrino absorption, it is important to discuss neutrino driven winds (e.g., \citet{Dessart:2008zd,Fujibayashi:2017puw}) in the context of the outflows found here.
A hot stellar remnant, such as the HMNS presented here, will likely produce strong mass outflows of the order of $\dot{M}\sim 10^{-3}-10^{-4}\, M_\odot /\rm yr$ \citep{Metzger:2018qfl}. This additional baryon loading could severely contaminate the environment, and lower the terminal Lorentz factors reached by the flares or steady-state outflows, potentially inhibiting their ability to power sGRBs.
On the other hand, if flares are indeed launched around the time of merger, they could potentially propagate ahead of the wind, resulting in a quasi-periodic precursor to  later sGRB production powered by a BH-disk system (in the case the HMNS collapses to a BH). Alternatively, since neutrino driven winds in the polar direction will be much slower than the outflow reported here, $v\sim 0.1-0.2 c$ \citep{Thompson:2001ys}, the relativistic outflow launched from the HMNS may be break out of the wind cloud, akin to jet breakout for the sGRB in GW170817 \citep{Gottlieb:2017pju,Lazzati:2017ocu,Xie:2018vya}.At the same time, the interaction with the wind may still affect the mass loading and terminal Lorentz factors of the outflows (see discussion above).

\subsection{Quasi-periodic oscillations}\label{sec:qpo}
Next we want to quantify the quasi-periodic substructure present in the electromagnetic outflows, and compare it to the frequency spectrum in the post-merger gravitational wave emission.
This will provide a first test of whether stellar oscillations of the HMNS can get imprinted in the electromagnetic outbursts, potentially driving sGRB emission at large scale.\\

{To this end}, we
show the frequency spectra in Fig. \ref{fig:freq}, and compare them to
post-merger gravitational wave frequency spectrum. Since the latter is in
part quasi-universally related to the dense matter equation of state
\citep{Bauswein:2012ya,Bauswein:2011tp,Takami:2014zpa,Bernuzzi:2015rla}, 
it has been suggested that QPOs in sGRBs could be used to provide
novel constraints \citep{Chirenti:2019sxw,Chirenti:2023dzl}. 
For both equations of state considered here, we
find that several frequencies are present in the Poynting luminosity.
In particular, corresponding to $1\, \rm kHz$, and at frequencies below
$500\, \rm Hz$. These modes are most dominant and correspond to the
periodic emission of flares. For a flare to inflate and detach roughly a
net rotation of $\Delta \phi = \pi$ is required. Assuming that this is
slowly provided by the net rotation difference of separate radial layers of
the HMNS, this is consistent with rotational profiles of HMNSs
reported in the literature (e.g., \citet{Hanauske:2016gia}).
Assuming a redshift of $z\simeq 0$, we find good agreement of the $1\, \rm
kHz$ peak in the strongly magnetized DD2 case with one of the GRB910711 frequencies \citep{Chirenti:2023dzl}.
We find that this agreement does not hold in the case of the APR$_{\rm LDP}$ equation of state.
At higher frequencies, we observe multiple distinct components at $\simeq
2\, \rm kHz$ and $3-4\, \rm kHz$, depending on the EOS used. These
frequencies lie within the expected band of the dominant $f_2$ frequency
peak present in the post-merger gravitational wave spectrum. Indeed,
for the strongly magnetized DD2 model we find good agreement between $f_2$ and a corresponding
frequency peak in the electromagnetic spectrum. However, this correlation
neither holds for the smaller magnetization, or the APR$_{\rm LDP}$ model, with the EM spectra appearing to have shifted to lower frequencies. Since both cases feature two peaks at higher frequencies, it is likely that
the frequencies should give rise to a non-linear coupling of the
frequencies.\\
However, in either case the power at high frequency is
suppressed by about two orders of magnitude, compared to the sub-kilohertz
part. Whether or not that would change potential imprints in sGRBs, with
the actual emission produced at much larger scales, remains unclear.

\section{Discussion}
Binary neutron star mergers can feature ultrastrong magnetic fields \citep{Kiuchi:2017zzg,Aguilera-Miret:2021fre}, produced by small-scale turbulent processes during merger \citep{Price:2006fi,Kiuchi:2015sga}. Since current numerical relativity simulations cannot capture all of these processes, the magnetic field topology and saturation strength after merger remains poorly constrained \citep{Aguilera-Miret:2020dhz,Palenzuela:2021gdo}.
One possibility is the generation of strong {toroidal} magnetic fields close to the surface of the star, where buoyant instabilities \citep{1978RSPTA.289..459A}  can then lead to a rise of poloidal loops into the surface layers \citep{Kiuchi:2011yt}. Strong differential rotation present in the HMNS \citep{Hanauske:2016gia,Kastaun:2016yaf} can then lead to inflation of these loops, and ultimately to their detachment and launching as flares \citep{Beloborodov:2013kpa}.\\
In this work, we have numerically investigated this scenario for launching such flares in the early post-merger phase of a binary neutron star coalescence. 
Using a dynamo sub-grid model to drive ultra-strong magnetic fields in the early stages of 
HMNS formation, we demonstrate that buoyant instabilities can give rise to quasi-periodic flaring from the surface of
the HMNS remnant. These flares bear some similarities with boosted fireball sGRB models \citep{Duffell:2013wra}.
Additional dissipation may be provided by internal kink instabilities inside the jet \citep{Bromberg:2015wra,Bromberg:2019pmh,Davelaar:2019mnr} .
The accompanying mildly relativistic collimated outflow features a quasi-periodic kilohertz substructure in the power spectrum of the outgoing Poynting flux. This behavior might potentially be relevant in the context of recently reported kilohertz QPOs in sGRBs \cite{Chirenti:2023dzl}.  The latter have been interpreted as being related to QPOs seen in gravitational wave spectra of HMNSs, which are caused by normal modes of the neutron star.   By contrast, we do not find  clear agreement between the frequency spectrum of the
electromagnetic outflow and the post-merger gravitational wave frequency spectrum.   
The former depends on the equation of state and magnetic field amplification, although we do find the same number of frequency peaks in both the electromagnetic and gravitational wave signals. 
More specifically, for the strongly magnetized simulation using the DD2 equation of state we obtain QPOs in the outgoing Poynting flux that are in good agreement
with the QPO frequencies reported for GRB910711 \citep{Chirenti:2023dzl}, when assuming a redshift
$z\simeq 0$.   This may, however, just be a coincidence since we do not see the same agreement for the APR EOS or the DD2 EOS and a weaker initial magnetic field.
{In all cases, the kilohertz components of the power spectrum are suppressed compared to the sub-kilohertz peak, which is directly associated with the flares and the relative velocity gradient at the surface of the star. Such a (dominant) sub-kHz QPO frequency has not been inferred from GRB910711.}
Further evidence for the origin of sub-kHz variability may also come from population statistics \citep{Camisasca:2023thc}.\\

While promising, our work points to the need to better understand the magnetic field topology sourced in the merger, 
and to clarify the launching mechanism of jet-like outflows from HMNSs \citep{Mosta:2020hlh}.  It is particularly important to understand how quickly very strong near-surface magnetic fields are created, and whether this depends on the EOS or merger properties (e.g., mass ratio), since this will influence whether buoyant instabilities can arise generically
on time scales of post-merger gravitational wave oscillations, or only on secular time scales when the remnant is already axisymmetric.
Beyond the topics highlighted in this work, the launching of early-time polar outflows is important because in principle they alone could power an sGRB.  The total Poynting flux in the outflows found here is much less than the maximal Poynting flux achievable with the $\sim 10^{17}$ G fields present in the HMNS.  However, it is possible that HMNS remnants that live longer, e.g., because of lower total mass in the merging system, can power significantly more energetic outflows that could be relevant to the bulk of the SGRB population.   The early time outflow driven by the HMNS could also be important if it clears the funnel region of matter.
This may have implications for further jet-launching from the system, e.g., if a black hole is formed at later times. 
It is also important to understand the impact of neutrino-driven winds \citep{Thompson:2001ys,Dessart:2008zd} on the mass-loading of the quasi-periodic flares found here. This will be crucial in determining whether the flares are in fact able to power (precursors to) sGRBs.   

\section*{Acknowledgments}
ERM thanks A. Beloborodov, K. Chatziioannou, A. Philippov, and C. Raithel for insightful discussions related to this
work. ERM gratefully acknowledges support as the John Archibald Wheeler Fellow at Princeton University (PCTS),
and from the Institute for Advanced Study. The simulations were performed on Delta at the National Center for Supercomputing Applications (NCSA) through allocation PHY210074 from the Advanced Cyberinfrastructure Coordination Ecosystem: Services \& Support (ACCESS) program, which is supported by National Science Foundation grants \#2138259, \#2138286, \#2138307, \#2137603, and \#2138296. 
Part of the simulations were carried out on NSF Frontera supercomputer under grant AST21006.   EQ was supported in part by a Simons Investigator award from the Simons Foundation.

\software{EinsteinToolkit \citep{Loffler:2011ay},
	  kuibit \citep{kuibit},
	  matplotlib \citep{Hunter:2007},
	  numpy \citep{harris2020array},
	  scipy \citep{2020SciPy-NMeth}
}

\bibliography{inspire,non_inspire}

\begin{thebibliography}{}
\expandafter\ifx\csname natexlab\endcsname\relax\def\natexlab#1{#1}\fi
\providecommand{\url}[1]{\href{#1}{#1}}
\providecommand{\dodoi}[1]{doi:~\href{http://doi.org/#1}{\nolinkurl{#1}}}
\providecommand{\doeprint}[1]{\href{http://ascl.net/#1}{\nolinkurl{http://ascl.net/#1}}}
\providecommand{\doarXiv}[1]{\href{https://arxiv.org/abs/#1}{\nolinkurl{https://arxiv.org/abs/#1}}}

\bibitem[{Abbott {et~al.}(2017)}]{LIGOScientific:2017zic}
Abbott, B.~P., {et~al.} 2017, Astrophys. J. Lett., 848, L13,
  \dodoi{10.3847/2041-8213/aa920c}

\bibitem[{{Acheson} \& {Gibbons}(1978)}]{1978RSPTA.289..459A}
{Acheson}, D.~J., \& {Gibbons}, M.~P. 1978, Philosophical Transactions of the
  Royal Society of London Series A, 289, 459, \dodoi{10.1098/rsta.1978.0066}

\bibitem[{Aguilera-Miret {et~al.}(2020)Aguilera-Miret, Vigan\`o, Carrasco,
  Mi\~nano, \& Palenzuela}]{Aguilera-Miret:2020dhz}
Aguilera-Miret, R., Vigan\`o, D., Carrasco, F., Mi\~nano, B., \& Palenzuela, C.
  2020, Phys. Rev. D, 102, 103006, \dodoi{10.1103/PhysRevD.102.103006}

\bibitem[{Aguilera-Miret {et~al.}(2022)Aguilera-Miret, Vigan\`o, \&
  Palenzuela}]{Aguilera-Miret:2021fre}
Aguilera-Miret, R., Vigan\`o, D., \& Palenzuela, C. 2022, Astrophys. J. Lett.,
  926, L31, \dodoi{10.3847/2041-8213/ac50a7}

\bibitem[{Baiotti \& Rezzolla(2017)}]{Baiotti:2016qnr}
Baiotti, L., \& Rezzolla, L. 2017, Rept. Prog. Phys., 80, 096901,
  \dodoi{10.1088/1361-6633/aa67bb}

\bibitem[{Bauswein \& Janka(2012)}]{Bauswein:2011tp}
Bauswein, A., \& Janka, H.~T. 2012, Phys. Rev. Lett., 108, 011101,
  \dodoi{10.1103/PhysRevLett.108.011101}

\bibitem[{Bauswein {et~al.}(2012)Bauswein, Janka, Hebeler, \&
  Schwenk}]{Bauswein:2012ya}
Bauswein, A., Janka, H.~T., Hebeler, K., \& Schwenk, A. 2012, Phys. Rev. D, 86,
  063001, \dodoi{10.1103/PhysRevD.86.063001}

\bibitem[{Beloborodov(2014)}]{Beloborodov:2013kpa}
Beloborodov, A.~M. 2014, Mon. Not. Roy. Astron. Soc., 438, 169,
  \dodoi{10.1093/mnras/stt2140}

\bibitem[{Berger(2014)}]{Berger:2013jza}
Berger, E. 2014, Ann. Rev. Astron. Astrophys., 52, 43,
  \dodoi{10.1146/annurev-astro-081913-035926}

\bibitem[{Bernuzzi {et~al.}(2015)Bernuzzi, Dietrich, \&
  Nagar}]{Bernuzzi:2015rla}
Bernuzzi, S., Dietrich, T., \& Nagar, A. 2015, Phys. Rev. Lett., 115, 091101,
  \dodoi{10.1103/PhysRevLett.115.091101}

\bibitem[{{Bhattacharjee} \& {Yuan}(1995)}]{1995ApJ...449..739B}
{Bhattacharjee}, A., \& {Yuan}, Y. 1995, \apj, 449, 739, \dodoi{10.1086/176094}

\bibitem[{{Bozzola}(2021)}]{kuibit}
{Bozzola}, G. 2021, The Journal of Open Source Software, 6, 3099,
  \dodoi{10.21105/joss.03099}

\bibitem[{Bozzola {et~al.}(2018)Bozzola, Stergioulas, \&
  Bauswein}]{Bozzola:2017qbu}
Bozzola, G., Stergioulas, N., \& Bauswein, A. 2018, Mon. Not. Roy. Astron.
  Soc., 474, 3557, \dodoi{10.1093/mnras/stx3002}

\bibitem[{Bromberg {et~al.}(2019)Bromberg, Singh, Davelaar, \&
  Philippov}]{Bromberg:2019pmh}
Bromberg, O., Singh, C.~B., Davelaar, J., \& Philippov, A.~A. 2019,
  \dodoi{10.3847/1538-4357/ab3fa5}

\bibitem[{Bromberg \& Tchekhovskoy(2016)}]{Bromberg:2015wra}
Bromberg, O., \& Tchekhovskoy, A. 2016, Mon. Not. Roy. Astron. Soc., 456, 1739,
  \dodoi{10.1093/mnras/stv2591}

\bibitem[{Bucciantini \& Del~Zanna(2013)}]{Bucciantini:2012sm}
Bucciantini, N., \& Del~Zanna, L. 2013, Mon. Not. Roy. Astron. Soc., 428, 71,
  \dodoi{10.1093/mnras/sts005}

\bibitem[{Camisasca {et~al.}(2023)}]{Camisasca:2023thc}
Camisasca, A.~E., {et~al.} 2023, Astron. Astrophys., 671, A112,
  \dodoi{10.1051/0004-6361/202245657}

\bibitem[{Carrasco \& Shibata(2020)}]{Carrasco:2020sxg}
Carrasco, F., \& Shibata, M. 2020, Phys. Rev. D, 101, 063017,
  \dodoi{10.1103/PhysRevD.101.063017}

\bibitem[{Carrasco {et~al.}(2019)Carrasco, Vigan\`o, Palenzuela, \&
  Pons}]{Carrasco:2019aas}
Carrasco, F., Vigan\`o, D., Palenzuela, C., \& Pons, J.~A. 2019, Mon. Not. Roy.
  Astron. Soc., 484, L124, \dodoi{10.1093/mnrasl/slz016}

\bibitem[{Chabanov {et~al.}(2023)Chabanov, Tootle, Most, \&
  Rezzolla}]{Chabanov:2022twz}
Chabanov, M., Tootle, S.~D., Most, E.~R., \& Rezzolla, L. 2023, Astrophys. J.
  Lett., 945, L14, \dodoi{10.3847/2041-8213/acbbc5}

\bibitem[{Chirenti {et~al.}(2023)Chirenti, Dichiara, Lien, Miller, \&
  Preece}]{Chirenti:2023dzl}
Chirenti, C., Dichiara, S., Lien, A., Miller, M.~C., \& Preece, R. 2023,
  Nature, 613, 253, \dodoi{10.1038/s41586-022-05497-0}

\bibitem[{Chirenti {et~al.}(2019)Chirenti, Miller, Strohmayer, \&
  Camp}]{Chirenti:2019sxw}
Chirenti, C., Miller, M.~C., Strohmayer, T., \& Camp, J. 2019, Astrophys. J.
  Lett., 884, L16, \dodoi{10.3847/2041-8213/ab43e0}

\bibitem[{Ciolfi(2020)}]{Ciolfi:2020hgg}
Ciolfi, R. 2020, Mon. Not. Roy. Astron. Soc., 495, L66,
  \dodoi{10.1093/mnrasl/slaa062}

\bibitem[{Ciolfi \& Kalinani(2020)}]{Ciolfi:2020wfx}
Ciolfi, R., \& Kalinani, J.~V. 2020, Astrophys. J. Lett., 900, L35,
  \dodoi{10.3847/2041-8213/abb240}

\bibitem[{Ciolfi {et~al.}(2017)Ciolfi, Kastaun, Giacomazzo, Endrizzi, Siegel,
  \& Perna}]{Ciolfi:2017uak}
Ciolfi, R., Kastaun, W., Giacomazzo, B., {et~al.} 2017, Phys. Rev. D, 95,
  063016, \dodoi{10.1103/PhysRevD.95.063016}

\bibitem[{Combi \& Siegel(2023)}]{Combi:2022nhg}
Combi, L., \& Siegel, D.~M. 2023, Astrophys. J., 944, 28,
  \dodoi{10.3847/1538-4357/acac29}

\bibitem[{Davelaar {et~al.}(2020)Davelaar, Philippov, Bromberg, \&
  Singh}]{Davelaar:2019mnr}
Davelaar, J., Philippov, A.~A., Bromberg, O., \& Singh, C.~B. 2020, Astrophys.
  J. Lett., 896, L31, \dodoi{10.3847/2041-8213/ab95a2}

\bibitem[{Del~Zanna {et~al.}(2007)Del~Zanna, Zanotti, Bucciantini, \&
  Londrillo}]{DelZanna:2007pk}
Del~Zanna, L., Zanotti, O., Bucciantini, N., \& Londrillo, P. 2007, Astron.
  Astrophys., 473, 11, \dodoi{10.1051/0004-6361:20077093}

\bibitem[{Dessart {et~al.}(2009)Dessart, Ott, Burrows, Rosswog, \&
  Livne}]{Dessart:2008zd}
Dessart, L., Ott, C.~D., Burrows, A., Rosswog, S., \& Livne, E. 2009,
  Astrophys. J., 690, 1681, \dodoi{10.1088/0004-637X/690/2/1681}

\bibitem[{Dionysopoulou {et~al.}(2015)Dionysopoulou, Alic, \&
  Rezzolla}]{Dionysopoulou:2015tda}
Dionysopoulou, K., Alic, D., \& Rezzolla, L. 2015, Phys. Rev. D, 92, 084064,
  \dodoi{10.1103/PhysRevD.92.084064}

\bibitem[{Duez {et~al.}(2005)Duez, Liu, Shapiro, \& Stephens}]{Duez:2005sf}
Duez, M.~D., Liu, Y.~T., Shapiro, S.~L., \& Stephens, B.~C. 2005, Phys. Rev. D,
  72, 024028, \dodoi{10.1103/PhysRevD.72.024028}

\bibitem[{Duffell \& MacFadyen(2013)}]{Duffell:2013wra}
Duffell, P.~C., \& MacFadyen, A.~I. 2013, Astrophys. J. Lett., 776, L9,
  \dodoi{10.1088/2041-8205/776/1/L9}

\bibitem[{Etienne {et~al.}(2015)Etienne, Paschalidis, Haas, M\"osta, \&
  Shapiro}]{Etienne:2015cea}
Etienne, Z.~B., Paschalidis, V., Haas, R., M\"osta, P., \& Shapiro, S.~L. 2015,
  Class. Quant. Grav., 32, 175009, \dodoi{10.1088/0264-9381/32/17/175009}

\bibitem[{Fern\'andez \& Metzger(2013)}]{Fernandez:2013tya}
Fern\'andez, R., \& Metzger, B.~D. 2013, Mon. Not. Roy. Astron. Soc., 435, 502,
  \dodoi{10.1093/mnras/stt1312}

\bibitem[{Fern\'andez {et~al.}(2019)Fern\'andez, Tchekhovskoy, Quataert,
  Foucart, \& Kasen}]{Fernandez:2018kax}
Fern\'andez, R., Tchekhovskoy, A., Quataert, E., Foucart, F., \& Kasen, D.
  2019, Mon. Not. Roy. Astron. Soc., 482, 3373, \dodoi{10.1093/mnras/sty2932}

\bibitem[{Fujibayashi {et~al.}(2018)Fujibayashi, Kiuchi, Nishimura, Sekiguchi,
  \& Shibata}]{Fujibayashi:2017puw}
Fujibayashi, S., Kiuchi, K., Nishimura, N., Sekiguchi, Y., \& Shibata, M. 2018,
  Astrophys. J., 860, 64, \dodoi{10.3847/1538-4357/aabafd}

\bibitem[{Giacomazzo {et~al.}(2011)Giacomazzo, Rezzolla, \&
  Baiotti}]{Giacomazzo:2010bx}
Giacomazzo, B., Rezzolla, L., \& Baiotti, L. 2011, Phys. Rev. D, 83, 044014,
  \dodoi{10.1103/PhysRevD.83.044014}

\bibitem[{Gill {et~al.}(2019)Gill, Nathanail, \& Rezzolla}]{Gill:2019bvq}
Gill, R., Nathanail, A., \& Rezzolla, L. 2019, Astrophys. J., 876, 139,
  \dodoi{10.3847/1538-4357/ab16da}

\bibitem[{Gottlieb {et~al.}(2018)Gottlieb, Nakar, Piran, \&
  Hotokezaka}]{Gottlieb:2017pju}
Gottlieb, O., Nakar, E., Piran, T., \& Hotokezaka, K. 2018, Mon. Not. Roy.
  Astron. Soc., 479, 588, \dodoi{10.1093/mnras/sty1462}

\bibitem[{Gressel \& Pessah(2015)}]{Gressel:2015mxa}
Gressel, O., \& Pessah, M.~E. 2015, Astrophys. J., 810, 59,
  \dodoi{10.1088/0004-637X/810/1/59}

\bibitem[{Gruzinov \& Diamond(1994)}]{Gruzinov:1994zz}
Gruzinov, A.~V., \& Diamond, P.~H. 1994, Phys. Rev. Lett., 72, 1651,
  \dodoi{10.1103/PhysRevLett.72.1651}

\bibitem[{Hanauske {et~al.}(2017)Hanauske, Takami, Bovard, Rezzolla, Font,
  Galeazzi, \& St\"ocker}]{Hanauske:2016gia}
Hanauske, M., Takami, K., Bovard, L., {et~al.} 2017, Phys. Rev. D, 96, 043004,
  \dodoi{10.1103/PhysRevD.96.043004}

\bibitem[{Hansen \& Lyutikov(2001)}]{Hansen:2000am}
Hansen, B. M.~S., \& Lyutikov, M. 2001, Mon. Not. Roy. Astron. Soc., 322, 695,
  \dodoi{10.1046/j.1365-8711.2001.04103.x}

\bibitem[{Harris {et~al.}(2020)Harris, Millman, van~der Walt, Gommers,
  Virtanen, Cournapeau, Wieser, Taylor, Berg, Smith, Kern, Picus, Hoyer, van
  Kerkwijk, Brett, Haldane, del R{\'{i}}o, Wiebe, Peterson,
  G{\'{e}}rard-Marchant, Sheppard, Reddy, Weckesser, Abbasi, Gohlke, \&
  Oliphant}]{harris2020array}
Harris, C.~R., Millman, K.~J., van~der Walt, S.~J., {et~al.} 2020, Nature, 585,
  357, \dodoi{10.1038/s41586-020-2649-2}

\bibitem[{Hempel \& Schaffner-Bielich(2010)}]{Hempel:2009mc}
Hempel, M., \& Schaffner-Bielich, J. 2010, Nucl. Phys. A, 837, 210,
  \dodoi{10.1016/j.nuclphysa.2010.02.010}

\bibitem[{Hilditch {et~al.}(2013)Hilditch, Bernuzzi, Thierfelder, Cao, Tichy,
  \& Bruegmann}]{Hilditch:2012fp}
Hilditch, D., Bernuzzi, S., Thierfelder, M., {et~al.} 2013, Phys. Rev. D, 88,
  084057, \dodoi{10.1103/PhysRevD.88.084057}

\bibitem[{Hogg \& Reynolds(2018)}]{Hogg:2018zon}
Hogg, J.~D., \& Reynolds, C. 2018, Astrophys. J., 861, 24,
  \dodoi{10.3847/1538-4357/aac439}

\bibitem[{Hotokezaka {et~al.}(2013)Hotokezaka, Kiuchi, Kyutoku, Muranushi,
  Sekiguchi, Shibata, \& Taniguchi}]{Hotokezaka:2013iia}
Hotokezaka, K., Kiuchi, K., Kyutoku, K., {et~al.} 2013, Phys. Rev. D, 88,
  044026, \dodoi{10.1103/PhysRevD.88.044026}

\bibitem[{Hotokezaka {et~al.}(2011)Hotokezaka, Kyutoku, Okawa, Shibata, \&
  Kiuchi}]{Hotokezaka:2011dh}
Hotokezaka, K., Kyutoku, K., Okawa, H., Shibata, M., \& Kiuchi, K. 2011, Phys.
  Rev. D, 83, 124008, \dodoi{10.1103/PhysRevD.83.124008}

\bibitem[{Hunter(2007)}]{Hunter:2007}
Hunter, J.~D. 2007, Computing in Science \& Engineering, 9, 90,
  \dodoi{10.1109/MCSE.2007.55}

\bibitem[{Ioka \& Taniguchi(2000)}]{Ioka:2000yb}
Ioka, K., \& Taniguchi, K. 2000, Astrophys. J., 537, 327,
  \dodoi{10.1086/309004}

\bibitem[{Kastaun {et~al.}(2016)Kastaun, Ciolfi, \&
  Giacomazzo}]{Kastaun:2016yaf}
Kastaun, W., Ciolfi, R., \& Giacomazzo, B. 2016, Phys. Rev. D, 94, 044060,
  \dodoi{10.1103/PhysRevD.94.044060}

\bibitem[{Kiuchi {et~al.}(2015)Kiuchi, Cerd\'a-Dur\'an, Kyutoku, Sekiguchi, \&
  Shibata}]{Kiuchi:2015sga}
Kiuchi, K., Cerd\'a-Dur\'an, P., Kyutoku, K., Sekiguchi, Y., \& Shibata, M.
  2015, Phys. Rev. D, 92, 124034, \dodoi{10.1103/PhysRevD.92.124034}

\bibitem[{Kiuchi {et~al.}(2022)Kiuchi, Fujibayashi, Hayashi, Kyutoku,
  Sekiguchi, \& Shibata}]{Kiuchi:2022nin}
Kiuchi, K., Fujibayashi, S., Hayashi, K., {et~al.} 2022.
\newblock \doarXiv{2211.07637}

\bibitem[{Kiuchi {et~al.}(2018)Kiuchi, Kyutoku, Sekiguchi, \&
  Shibata}]{Kiuchi:2017zzg}
Kiuchi, K., Kyutoku, K., Sekiguchi, Y., \& Shibata, M. 2018, Phys. Rev. D, 97,
  124039, \dodoi{10.1103/PhysRevD.97.124039}

\bibitem[{Kiuchi {et~al.}(2014)Kiuchi, Kyutoku, Sekiguchi, Shibata, \&
  Wada}]{Kiuchi:2014hja}
Kiuchi, K., Kyutoku, K., Sekiguchi, Y., Shibata, M., \& Wada, T. 2014, Phys.
  Rev. D, 90, 041502, \dodoi{10.1103/PhysRevD.90.041502}

\bibitem[{Kiuchi {et~al.}(2011)Kiuchi, Yoshida, \& Shibata}]{Kiuchi:2011yt}
Kiuchi, K., Yoshida, S., \& Shibata, M. 2011, Astron. Astrophys., 532, A30,
  \dodoi{10.1051/0004-6361/201016242}

\bibitem[{Lazzati {et~al.}(2017)Lazzati, L\'opez-C\'amara, Cantiello, Ciolfi,
  Giacomazzo, Morsony, Perna, \& Workman}]{Lazzati:2017ocu}
Lazzati, D., L\'opez-C\'amara, D., Cantiello, M., {et~al.} 2017, Astrophys. J.
  Lett., 848, L6, \dodoi{10.3847/2041-8213/aa8f3d}

\bibitem[{Liu {et~al.}(2022)Liu, Kretschmer, \& Teyssier}]{liu2022subgrid}
Liu, Y., Kretschmer, M., \& Teyssier, R. 2022, Monthly Notices of the Royal
  Astronomical Society, 513, 6028

\bibitem[{Liu {et~al.}(2008)Liu, Shapiro, Etienne, \& Taniguchi}]{Liu:2008xy}
Liu, Y.~T., Shapiro, S.~L., Etienne, Z.~B., \& Taniguchi, K. 2008, Phys. Rev.
  D, 78, 024012, \dodoi{10.1103/PhysRevD.78.024012}

\bibitem[{Loffler {et~al.}(2012)}]{Loffler:2011ay}
Loffler, F., {et~al.} 2012, Class. Quant. Grav., 29, 115001,
  \dodoi{10.1088/0264-9381/29/11/115001}

\bibitem[{Mahlmann {et~al.}(2023)Mahlmann, Philippov, Mewes, Ripperda, Most, \&
  Sironi}]{Mahlmann:2023ipm}
Mahlmann, J.~F., Philippov, A.~A., Mewes, V., {et~al.} 2023.
\newblock \doarXiv{2302.07273}

\bibitem[{Margalit {et~al.}(2022)Margalit, Jermyn, Metzger, Roberts, \&
  Quataert}]{Margalit:2022rde}
Margalit, B., Jermyn, A.~S., Metzger, B.~D., Roberts, L.~F., \& Quataert, E.
  2022, Astrophys. J., 939, 51, \dodoi{10.3847/1538-4357/ac8b01}

\bibitem[{Martin {et~al.}(2015)Martin, Perego, Arcones, Thielemann, Korobkin,
  \& Rosswog}]{Martin:2015hxa}
Martin, D., Perego, A., Arcones, A., {et~al.} 2015, Astrophys. J., 813, 2,
  \dodoi{10.1088/0004-637X/813/1/2}

\bibitem[{Metzger(2020)}]{Metzger:2019zeh}
Metzger, B.~D. 2020, Living Rev. Rel., 23, 1, \dodoi{10.1007/s41114-019-0024-0}

\bibitem[{Metzger {et~al.}(2018)Metzger, Thompson, \&
  Quataert}]{Metzger:2018qfl}
Metzger, B.~D., Thompson, T.~A., \& Quataert, E. 2018, Astrophys. J., 856, 101,
  \dodoi{10.3847/1538-4357/aab095}

\bibitem[{Most {et~al.}(2019)Most, Papenfort, \& Rezzolla}]{Most:2019kfe}
Most, E.~R., Papenfort, L.~J., \& Rezzolla, L. 2019, Mon. Not. Roy. Astron.
  Soc., 490, 3588, \dodoi{10.1093/mnras/stz2809}

\bibitem[{Most \& Philippov(2020)}]{Most:2020ami}
Most, E.~R., \& Philippov, A.~A. 2020, Astrophys. J. Lett., 893, L6,
  \dodoi{10.3847/2041-8213/ab8196}

\bibitem[{Most \& Philippov(2022{\natexlab{a}})}]{Most:2022ayk}
---. 2022{\natexlab{a}}.
\newblock \doarXiv{2207.14435}

\bibitem[{Most \& Philippov(2022{\natexlab{b}})}]{Most:2022ojl}
---. 2022{\natexlab{b}}, Mon. Not. Roy. Astron. Soc., 515, 2710,
  \dodoi{10.1093/mnras/stac1909}

\bibitem[{Most \& Raithel(2021)}]{Most:2021ktk}
Most, E.~R., \& Raithel, C.~A. 2021, Phys. Rev. D, 104, 124012,
  \dodoi{10.1103/PhysRevD.104.124012}

\bibitem[{Most {et~al.}(2021)Most, Harris, Plumberg, Alford, Noronha,
  Noronha-Hostler, Pretorius, Witek, \& Yunes}]{Most:2021zvc}
Most, E.~R., Harris, S.~P., Plumberg, C., {et~al.} 2021, Mon. Not. Roy. Astron.
  Soc., 509, 1096, \dodoi{10.1093/mnras/stab2793}

\bibitem[{M\"osta {et~al.}(2020)M\"osta, Radice, Haas, Schnetter, \&
  Bernuzzi}]{Mosta:2020hlh}
M\"osta, P., Radice, D., Haas, R., Schnetter, E., \& Bernuzzi, S. 2020,
  Astrophys. J. Lett., 901, L37, \dodoi{10.3847/2041-8213/abb6ef}

\bibitem[{Nedora {et~al.}(2021)Nedora, Radice, Bernuzzi, Perego, Daszuta,
  Endrizzi, Prakash, \& Schianchi}]{Nedora:2021eoj}
Nedora, V., Radice, D., Bernuzzi, S., {et~al.} 2021, Mon. Not. Roy. Astron.
  Soc., 506, 5908, \dodoi{10.1093/mnras/stab2004}

\bibitem[{\"Ozel \& Freire(2016)}]{Ozel:2016oaf}
\"Ozel, F., \& Freire, P. 2016, Ann. Rev. Astron. Astrophys., 54, 401,
  \dodoi{10.1146/annurev-astro-081915-023322}

\bibitem[{Palenzuela {et~al.}(2022)Palenzuela, Aguilera-Miret, Carrasco,
  Ciolfi, Kalinani, Kastaun, Mi\~nano, \& Vigan\`o}]{Palenzuela:2021gdo}
Palenzuela, C., Aguilera-Miret, R., Carrasco, F., {et~al.} 2022, Phys. Rev. D,
  106, 023013, \dodoi{10.1103/PhysRevD.106.023013}

\bibitem[{Palenzuela {et~al.}(2013)Palenzuela, Lehner, Ponce, Liebling,
  Anderson, Neilsen, \& Motl}]{Palenzuela:2013hu}
Palenzuela, C., Lehner, L., Ponce, M., {et~al.} 2013, Phys. Rev. Lett., 111,
  061105, \dodoi{10.1103/PhysRevLett.111.061105}

\bibitem[{Palenzuela {et~al.}(2015)Palenzuela, Liebling, Neilsen, Lehner,
  Caballero, O'Connor, \& Anderson}]{Palenzuela:2015dqa}
Palenzuela, C., Liebling, S.~L., Neilsen, D., {et~al.} 2015, Phys. Rev. D, 92,
  044045, \dodoi{10.1103/PhysRevD.92.044045}

\bibitem[{Papenfort {et~al.}(2021)Papenfort, Tootle, Grandcl\'ement, Most, \&
  Rezzolla}]{Papenfort:2021hod}
Papenfort, L.~J., Tootle, S.~D., Grandcl\'ement, P., Most, E.~R., \& Rezzolla,
  L. 2021, Phys. Rev. D, 104, 024057, \dodoi{10.1103/PhysRevD.104.024057}

\bibitem[{Parfrey {et~al.}(2013)Parfrey, Beloborodov, \& Hui}]{Parfrey:2013gza}
Parfrey, K., Beloborodov, A.~M., \& Hui, L. 2013, Astrophys. J., 774, 92,
  \dodoi{10.1088/0004-637X/774/2/92}

\bibitem[{Poudel {et~al.}(2020)Poudel, Tichy, Br\"ugmann, \&
  Dietrich}]{Poudel:2020fte}
Poudel, A., Tichy, W., Br\"ugmann, B., \& Dietrich, T. 2020, Phys. Rev. D, 102,
  104014, \dodoi{10.1103/PhysRevD.102.104014}

\bibitem[{Price \& Rosswog(2006)}]{Price:2006fi}
Price, D., \& Rosswog, S. 2006, Science, 312, 719,
  \dodoi{10.1126/science.1125201}

\bibitem[{Raithel \& Most(2022)}]{Raithel:2022orm}
Raithel, C.~A., \& Most, E.~R. 2022, Astrophys. J. Lett., 933, L39,
  \dodoi{10.3847/2041-8213/ac7c75}

\bibitem[{Rezzolla {et~al.}(2011)Rezzolla, Giacomazzo, Baiotti, Granot,
  Kouveliotou, \& Aloy}]{Rezzolla:2011da}
Rezzolla, L., Giacomazzo, B., Baiotti, L., {et~al.} 2011, Astrophys. J. Lett.,
  732, L6, \dodoi{10.1088/2041-8205/732/1/L6}

\bibitem[{Ruffert {et~al.}(1996)Ruffert, Janka, \& Schaefer}]{Ruffert:1995fs}
Ruffert, M.~H., Janka, H.~T., \& Schaefer, G. 1996, Astron. Astrophys., 311,
  532.
\newblock \doarXiv{astro-ph/9509006}

\bibitem[{Ruiz {et~al.}(2016)Ruiz, Lang, Paschalidis, \&
  Shapiro}]{Ruiz:2016rai}
Ruiz, M., Lang, R.~N., Paschalidis, V., \& Shapiro, S.~L. 2016, Astrophys. J.
  Lett., 824, L6, \dodoi{10.3847/2041-8205/824/1/L6}

\bibitem[{Schneider {et~al.}(2019)Schneider, Constantinou, Muccioli, \&
  Prakash}]{Schneider:2019vdm}
Schneider, A.~S., Constantinou, C., Muccioli, B., \& Prakash, M. 2019, Phys.
  Rev. C, 100, 025803, \dodoi{10.1103/PhysRevC.100.025803}

\bibitem[{Shapiro(2000)}]{Shapiro:2000zh}
Shapiro, S.~L. 2000, Astrophys. J., 544, 397, \dodoi{10.1086/317209}

\bibitem[{Sharma {et~al.}(2023)Sharma, Barkov, \& Lyutikov}]{Sharma:2023cxc}
Sharma, P., Barkov, M., \& Lyutikov, M. 2023.
\newblock \doarXiv{2302.08848}

\bibitem[{Shibata {et~al.}(2021)Shibata, Fujibayashi, \&
  Sekiguchi}]{Shibata:2021xmo}
Shibata, M., Fujibayashi, S., \& Sekiguchi, Y. 2021, Phys. Rev. D, 104, 063026,
  \dodoi{10.1103/PhysRevD.104.063026}

\bibitem[{Shibata {et~al.}(2011)Shibata, Suwa, Kiuchi, \&
  Ioka}]{Shibata:2011fj}
Shibata, M., Suwa, Y., Kiuchi, K., \& Ioka, K. 2011, Astrophys. J. Lett., 734,
  L36, \dodoi{10.1088/2041-8205/734/2/L36}

\bibitem[{Siegel {et~al.}(2014)Siegel, Ciolfi, \& Rezzolla}]{Siegel:2014ita}
Siegel, D.~M., Ciolfi, R., \& Rezzolla, L. 2014, Astrophys. J. Lett., 785, L6,
  \dodoi{10.1088/2041-8205/785/1/L6}

\bibitem[{Siegel \& Metzger(2017)}]{Siegel:2017nub}
Siegel, D.~M., \& Metzger, B.~D. 2017, Phys. Rev. Lett., 119, 231102,
  \dodoi{10.1103/PhysRevLett.119.231102}

\bibitem[{Skoutnev {et~al.}(2021)Skoutnev, Most, Bhattacharjee, \&
  Philippov}]{Skoutnev:2021chg}
Skoutnev, V., Most, E.~R., Bhattacharjee, A., \& Philippov, A.~A. 2021,
  Astrophys. J., 921, 75, \dodoi{10.3847/1538-4357/ac1ba4}

\bibitem[{Stergioulas {et~al.}(2011)Stergioulas, Bauswein, Zagkouris, \&
  Janka}]{Stergioulas:2011gd}
Stergioulas, N., Bauswein, A., Zagkouris, K., \& Janka, H.-T. 2011, Mon. Not.
  Roy. Astron. Soc., 418, 427, \dodoi{10.1111/j.1365-2966.2011.19493.x}

\bibitem[{Takami {et~al.}(2014)Takami, Rezzolla, \& Baiotti}]{Takami:2014zpa}
Takami, K., Rezzolla, L., \& Baiotti, L. 2014, Phys. Rev. Lett., 113, 091104,
  \dodoi{10.1103/PhysRevLett.113.091104}

\bibitem[{Thompson {et~al.}(2001)Thompson, Burrows, \& Meyer}]{Thompson:2001ys}
Thompson, T.~A., Burrows, A., \& Meyer, B.~S. 2001, Astrophys. J., 562, 887,
  \dodoi{10.1086/323861}

\bibitem[{Tomei {et~al.}(2020)Tomei, Del~Zanna, Bugli, \&
  Bucciantini}]{Tomei:2019zpj}
Tomei, N., Del~Zanna, L., Bugli, M., \& Bucciantini, N. 2020, Mon. Not. Roy.
  Astron. Soc., 491, 2346, \dodoi{10.1093/mnras/stz3146}

\bibitem[{Tsang {et~al.}(2012)Tsang, Read, Hinderer, Piro, \&
  Bondarescu}]{Tsang:2011ad}
Tsang, D., Read, J.~S., Hinderer, T., Piro, A.~L., \& Bondarescu, R. 2012,
  Phys. Rev. Lett., 108, 011102, \dodoi{10.1103/PhysRevLett.108.011102}

\bibitem[{Virtanen {et~al.}(2020)Virtanen, Gommers, Oliphant, Haberland, Reddy,
  Cournapeau, Burovski, Peterson, Weckesser, Bright, {van der Walt}, Brett,
  Wilson, Millman, Mayorov, Nelson, Jones, Kern, Larson, Carey, Polat, Feng,
  Moore, {VanderPlas}, Laxalde, Perktold, Cimrman, Henriksen, Quintero, Harris,
  Archibald, Ribeiro, Pedregosa, {van Mulbregt}, \& {SciPy 1.0
  Contributors}}]{2020SciPy-NMeth}
Virtanen, P., Gommers, R., Oliphant, T.~E., {et~al.} 2020, Nature Methods, 17,
  261, \dodoi{10.1038/s41592-019-0686-2}

\bibitem[{Weih {et~al.}(2018)Weih, Most, \& Rezzolla}]{Weih:2017mcw}
Weih, L.~R., Most, E.~R., \& Rezzolla, L. 2018, Mon. Not. Roy. Astron. Soc.,
  473, L126, \dodoi{10.1093/mnrasl/slx178}

\bibitem[{Werneck {et~al.}(2022)}]{Werneck:2022exo}
Werneck, L.~R., {et~al.} 2022.
\newblock \doarXiv{2208.14487}

\bibitem[{White {et~al.}(2022)White, Burrows, Coleman, \&
  Vartanyan}]{White:2021erz}
White, C.~J., Burrows, A., Coleman, M. S.~B., \& Vartanyan, D. 2022, Astrophys.
  J., 926, 111, \dodoi{10.3847/1538-4357/ac4507}

\bibitem[{Xie {et~al.}(2018)Xie, Zrake, \& MacFadyen}]{Xie:2018vya}
Xie, X., Zrake, J., \& MacFadyen, A. 2018, Astrophys. J., 863, 58,
  \dodoi{10.3847/1538-4357/aacf9c}

\bibitem[{Zhang(2020)}]{Zhang:2020eou}
Zhang, B. 2020, Astrophys. J. Lett., 890, L24, \dodoi{10.3847/2041-8213/ab7244}

\bibitem[{Zhu {et~al.}(2020)Zhu, Li, \& Rezzolla}]{Zhu:2020imp}
Zhu, Z., Li, A., \& Rezzolla, L. 2020, Phys. Rev. D, 102, 084058,
  \dodoi{10.1103/PhysRevD.102.084058}

\end{thebibliography}
\end{document}